# Nitrogen Cluster Anions

Nikolaus Weinberger,[a] Johannes Postler,[a] Paul Scheier,[a,c] Olof Echt[a,b,c]

[a] *Institut für Ionenphysik und Angewandte Physik, University of Innsbruck, Technikerstrasse 25, A-6020 Innsbruck, Austria*
[b] *Department of Physics, University of New Hampshire, Durham, New Hampshire 03824, United States*

[c] Corresponding Authors
P. Scheier. E-mail: paul.scheier@uibk.ac.at. Phone: +43 512 507 52660. Fax: +43 512 507 2922.
O. Echt. E-mail: olof.echt@unh.edu. Phone: +1 603 862 3548. Fax: +01 603 862 2998.

Email addresses of other authors:
Postler, Johannes Johannes.Postler@uibk.ac.at

**Abstract**
Anions are formed by electron attachment to helium nanodroplets doped with $N_2$. The most prominent ion series is due to odd-numbered $N_m^-$ with $3 \leq m < 140$. Neither $N^-$ nor $N_2^-$ are observed. An appearance energy of $11.5 \pm 0.5$ eV is measured for $N_3^-$. The yield of $N_m^-$ averaged over $5 \leq m \leq 21$ shows an appearance energy of 9 eV, just above the estimated thermodynamic threshold for formation of $N_3^-$ from electron attachment to small $N_2$ clusters embedded in helium droplets. These findings support the notion that odd-numbered $N_m^-$ cluster ions are best characterized as $N_2$ van der Waals clusters with an azide anion chromophore but they are at odds with some theoretical reports. $N_3^-(N_2)_4$ and $N_3^-(N_2)_{11}$ form local maxima in the abundance distribution, suggesting that these ions are particularly stable. The yield of even-numbered $N_m^-$ ions ($m \geq 4$) is two orders of magnitude lower, barely exceeding the background level. Several other cluster anion series are observed that involve impurities.

## 1. Introduction

Recent interest in polynitrogen molecules is motivated by the search for clean-energy high-density materials. The triple bond in $N_2$ is much stronger than the sum of three single bonds, therefore decomposition of single-bonded polynitrogen into $N_2$ would release enormous amounts of energy without harmful waste.[1-2] Several polymeric neutral and charged complexes have been investigated (for a review, see Samartzis and Wodtke[1]); Sharma et al. have suggested that polymeric nitrogen may be formed by confinement, for example in a fullerene cage.[3]

Numerous reports have been devoted to positively charged nitrogen clusters which are efficiently generated by electron ionization of neutral van der Waals clusters formed in supersonic jets,[4-10] condensation on seed ions in gases,[11-12] sputtering of frozen $N_2$ by energetic atoms or ions,[13-17] or by doping helium nanodroplets with $N_2$ and subsequent electron ionization.[18]

On the other hand, there are very few reports on negatively charged nitrogen clusters. An unusual feature of $N_2$ is that neither the intact molecule nor its dissociation products form stable anions (although electronically excited metastable anion states have been observed[19-20]). The next larger all-nitrogen molecule, $N_3$, is metastable with respect to dissociation into $N_2 + N$[21] but the azide anion is stable, with a detachment energy of $2.68 \pm 0.03$ eV.[22]

Reports on all-nitrogen molecular anions larger than $N_3^-$ are scarce. Sputtering of frozen $O_2$ and other molecular gases such as $H_2O$ or $CO_2$ by fission fragments from $^{252}Cf$ (which emits heavy ions such as $Ba^{15+}$ with kinetic energies of about 65 MeV) leads to emission of various cluster anions but no $N_m^-$ anions have been observed in these experiments upon sputtering of frozen $N_2$.[23-24] Jonkman and Michl have sputtered frozen $N_2$ with 4 keV $Ar^+$ and observed copious amounts of odd and even-numbered $N_m^+$ cations but no anions.[13] On the other hand, Tonuma et al. have bombarded frozen nitrogen with $Ar^{13+}$ accelerated to 60 MeV; they report formation of odd-numbered $N_m^-$ with $1 \le m \le 13$ and even-numbered $N_m^-$ with $2 \le m \le 8$.[16] The intensities of anions which were dominated by $N_3^-$ was extremely weak, about 4 to 5 orders of magnitude less than those of cations. Havel and coworkers reported the appearance of $N_2^-$ and $N_3^-$ upon laser desorption of aluminum nitride; $N_6^-$ and $N_{10}^-$ through $N_{15}^-$ were observed upon laser desorption of phosphorous nitride.[25-26]

The only other report on nitrogen cluster anions that we are aware of is by Vostrikov and Dubov who collided $N_2$ clusters formed in a supersonic jet with electrons.[27] The expansion conditions were varied to produce $N_2$ clusters with average sizes ranging from $(N_2)_{125}$ to $(N_2)_{520}$; the electron energy was varied from 0 to 35 eV. They observed a pronounced resonance in the anion yield at about 2 eV which was assumed to be due to $(N_2)_n^-$ but no mass spectra were recorded.

Here we report the formation of nitrogen cluster anions by electron attachment to helium nanodroplets that were doped with molecular nitrogen, combined with mass spectral analysis. Odd numbered $N_m^-$ ($m \ge 3$) are the most abundant ions, with $m$ ranging beyond 140. A measurement of the ion yield *versus* electron energy suggests that these ions are best characterized as $N_3^-(N_2)_n$. A prominent anomaly in the abundance distribution suggests that $N_3^-(N_2)_{11}$ is particularly stable. Given the linear structure of $N_3^-$ it is not easy to surmise the structure of $N_3^-(N_2)_{11}$. Even-numbered $N_m^-$ ($m \ge 4$) barely exceed the background level. Another ion with enhanced abundance is $N_3^-(N_2)_4$. Several other ion series are observed which involve oxygen, water or carbon impurities.

## 2. Experimental

Helium nanodroplets were produced by expanding helium (Messer, purity 99.9999 %) at a stagnation pressure of 26 bar through a 5 μm nozzle, cooled by a closed-cycle refrigerator to 10.0 K, into a vacuum chamber (base pressure about $2\times10^{-6}$ Pa). Droplets that form in the (continuous) expansion contain an average number of $10^5$ atoms; they are superfluid with a temperature of 0.37 K.[28] The resulting supersonic beam was skimmed by a 0.8 mm conical skimmer and traversed a 20 cm long pick-up cell into which nitrogen gas (Messer, purity 99.999 %) was introduced. The nitrogen pressure was varied in order to obtain the optimal experimental conditions for formation of nitrogen cluster ions.

The beam of doped helium droplets was collimated and crossed by an electron beam in a Nier-type ion source. Ions were accelerated into the extraction region of a reflectron time-of-flight mass spectrometer (Tofwerk AG, model HTOF) with a mass resolution $\Delta m/m = 1/3800$ ($\Delta m$ = full-width-at-half-maximum). The base pressure in the mass spectrometer was $10^{-5}$ Pa. Ions were extracted at 90° into the field-free region of the spectrometer by a pulsed voltage. At the end of the field-free region they entered a two-stage reflectron which reflected them towards a microchannel plate detector operated in single ion counting mode.

The acquisition time of mass spectra presented here ranged from 4.3 to 10.3 hours. During acquisition the electron energy which could be tuned between 1.5 and 50.5 eV was periodically scanned in increments of 0.05 eV. Spectra were stored every minute or so, allowing to evaluate the ion yield of specific ions as a function of electron energy. In order to reduce statistical scatter, mass spectra shown in this work comprise data accumulated over a wide range of electron energies (several eV) which will be specified in the result section.

Mass spectra were evaluated by means of a custom-designed software.[29] The routine includes automatic fitting of a custom peak shape to the mass peaks and subtraction of background by fitting a spline to the background level of the raw data. The abundance of ions with a specific composition (specific values of $m$ and $n$) is derived by a matrix method. Additional experimental details have been provided elsewhere.[30-31]

## 3. Results

A mass spectrum of negative ions formed by electron attachment to helium nanodroplets doped with nitrogen is shown in Fig. 1a. The electron energy was repeatedly scanned from 19.5 to 50.5 eV as explained in the Experimental Section. The most prominent ions have the composition $N_m^-$ with odd $m \geq 3$. There is no evidence for $N^-$. These ions are labeled $N_3^-(N_2)_n$ because, presumably, the azide anion forms the ionic core as discussed further below. Some values of $n$ are indicated in Fig. 1a.

Two other prominent ion series that contain impurities are visible in Fig. 1a. Their mass peaks are connected by red and blue lines for the stronger and weaker series, respectively. These and other impurity ions are identified in Fig. 1b which zooms into the mass region between $N_3^-(N_2)_7$ and $N_3^-(N_2)_8$. Their likely composition, listed in the order of increasing mass, is $O^-(N_2)_8$ at nominally 240 u, $N_3^-(N_2)_7He$ at 242 u, $O_2^-(N_2)_7H_2O$ and $N_3^-(N_2)_6(H_2O)_2$ at 246 u, $C_2N_2^-(N_2)_7$ at 248 u, $CN^-(N_2)_8$, at 250 u, and $O_2^-(N_2)_8$ and $N_3^-(N_2)_7H_2O$ at 256 u. The likely nature of the chromophores in these ions will be discussed in Section 4. Pairs of isobaric ions are listed because $N+H_2O$ has the same nominal mass as $O_2$; the former is 0.024 u heavier than the latter. The mass peaks are not resolved but a fit indicates that both ions contribute; the ion containing an extra $H_2O$ tends to be more abundant.

Three mass peaks in Fig. 1b, flagged by short vertical arrows, occur to the right of the three most prominent mass peaks. They are assigned to isotopologues of $N_3^-(N_2)_7$, $O^-(N_2)_8$, and $O_2^-(N_2)_8$ and $N_3^-(N_2)_7H_2O$ that contain one $^{15}N$ whose natural abundance is 0.386 %. The expected abundance ratio of $^{14}N_{17}$ versus $^{15}N^{14}N_{16}$ is 15:1, in good agreement with the spectrum. Other isotopologues (ions containing $^{13}C$ or $^{18}O$) would be buried in the background.

The mass spectrum in Fig. 1a suggests that the ion yield does not always vary smoothly with size. The most prominent anomalies occur for the ion series $N_3^-(N_2)_n$ and $C_2N_2^-(N_2)_n$. Their abundances are extracted from the mass spectrum by means of custom-designed software.[29] The routine includes automatic fitting of mass peaks and subtraction of background; it explicitly considers isotopic patterns of all ions. The abundance of specific ions (specific value of $n$) is derived by a matrix method. The results are displayed in Fig. 2. The $N_3^-(N_2)_n$ series (Fig. 2a) exhibits an abrupt drop from $n = 11$ to 12 and a weaker one from $n = 4$ to 5. The distribution of $C_2N_2^-(N_2)_n$ (Fig. 2b) exhibits a distinct drop beyond $n = 15$. Distributions extracted from two other mass spectra, recorded at electron energies below 20.5 eV, show similar features.

We have searched for even-numbered $N_m^-$ anions by fitting Gaussians of predetermined position and width. They seem to exist for $m \geq 4$ but their yield is typically $1.0 \pm 0.3$ % of the yield of the nearest odd-numbered nitrogen cluster ions. For example, in Fig. 1b $(N_2)_9^-$ is present at 252.055 u with an amplitude three times less than that of $N_3^-(N_2)_7He$, the weakest impurity ion. There is no evidence for $N_2^-$.

The spectrum displayed in Fig. 1 was recorded for a duration of 10.3 hours. Two other spectra were recorded, one with the energy scanned from 10.5 to 20.5 eV and an acquisition time of 5 hours, and another one with the energy scanned from 1.4 to 13.7 eV (acquisition time 4.2 hours). These spectra have poorer statistics than the one shown in Fig. 1. The spectrum recorded at 1.4 to 13.7 eV, in particular, shows a much lower ion yield; no ions other than odd-numbered $N_m^-$ can be identified. However, data from these three spectra have been combined to extract the energy dependence for formation of anions; the result is shown in Fig. 3. The yield of $N_3^-$ (Fig. 3a) is negligible below 10 eV; it rises steeply from 10 to 20 eV and decreases beyond 27 eV. The energy threshold $E_0$ is extracted by a non-linear least squares fit of a Wannier-type power law (with a fitted exponent of 1.283) shown in Fig. 3a as a solid line, taking into account the finite energy resolution of 1.5 eV (FWHM) by analytically convoluting the power law with a Gaussian before the numerical fitting procedure.[32] The result is $E_0 = 11.5 \pm 0.5$ eV. The energy dependence of $N_3^-(N_2)_n$ ions, averaged over $1 \leq n \leq 9$, is displayed in Fig. 3b. The overall shape is similar to that of $N_3^-$ but there is a distinct stepwise onset at about 9 eV.

## 4. Discussion
### 4.1 Pure $N_m^-$ anions

The most abundant anions observed in the present work have the composition $N_m^-$ where $m \geq 3$, odd. The abundance of even-numbered $N_m^-$ ions ($m \geq 4$) is about $1.0 \pm 0.3$ % of adjacent odd-numbered ions, barely above background. What is the chromophore in the odd-numbered ions? Atomic nitrogen has a negative electron affinity.[20] Hund's rules predict the lowest $N^-$ state to be $2p^4\ ^3P$; scattering experiments indicate that this state is 0.07 eV above a free electron plus N in its $2p^3\ ^4S$ ground state, in rough agreement with various calculated values.[20, 33] Hiraoka et al. have detected metastable $N^-$ in mass spectra by electron excitation of gaseous NO or a mixture of $N_2 + O_2$.[34] An appearance energy of 16 eV was found for the $N_2 + O_2$ mixture. The authors estimate that the thermodynamic threshold of $N^-$ ($^1D$) formed by dissociative attachment to $N_2$ must be greater than 13.58 eV.

Fig. 3 shows an appearance energy of $11.5 \pm 0.5$ eV for $N_3^-$. When comparing this value with Hiraoka's work one needs to consider the effect of the helium environment on the energetics. The bottom of the electron conduction band in helium lies above the vacuum level.[35] As a result, molecules embedded in helium droplets feature resonances in the anion yield that are blue shifted by about 1.6 eV relative to resonances observed for bare molecules.[36-37] Thus, the observed appearance energy of 11.5 eV is much less than the threshold of $13.58 + 1.6 = 15.18$ eV expected if the reaction were to proceed via formation of metastable $N^-$ ($^1D$) in a helium nanodroplet.

A more likely chromophore in odd-numbered $N_m^-$ is the azide anion. The $N_3$ radical is linear with $D_{h\infty}$ symmetry.[1] It is metastable with respect to dissociation into $N + N_2$ with $D_0 = -0.05 \pm 0.10$ eV.[21] Its large adiabatic electron affinity of $2.68 \pm 0.03$ eV renders a stable anion that is linear with centrosymmetric charge distribution.[22] Combining these values with the dissociation energies of $N_2$ (9.904 eV, ref. [38] and references therein) and the $(N_2)_2$ van der Waals dimer (13 meV[39]) one obtains the energy threshold for $N_3^-$ formation from bare $(N_2)_2$,

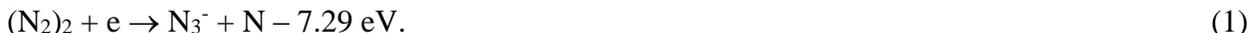
$(N_2)_2 + e \rightarrow N_3^- + N - 7.29$ eV. (1)

For $(N_2)_2$ embedded in helium droplets the expected thermodynamic threshold for $N_3^-$ formation would be shifted upward by about 1.6 eV to 8.89 eV. This is nearly 2.6 eV below the appearance energy of $N_3^-$ (11.5 eV, see Fig. 3a), but very close to the 9 eV appearance energy of $N_3^-(N_2)_n$, averaged over $1 \leq n \leq 9$ (Fig. 3b). The ultracold (0.37 K) helium droplet ensures the absence of thermal excitation in the neutral precursor molecule, hence appearance energies will present upper bounds to thermodynamic values. The bond length of $N_2$, 1.098 Å,[40] is significantly shorter than the calculated NN bond length in $N_3^-$ of 1.18 Å.[22] On the other hand, the NN separation between atoms in two distinct $N_2$ molecules in a nitrogen van der Waals cluster will be much larger than 1.18 Å. The mismatch explains why the appearance energy of $N_3^-$ significantly exceeds the thermodynamic value, but the droplet may help to approach the thermodynamic threshold when cluster ions $N_3^-(N_2)_n$ are probed.

Theoretical studies of odd-numbered $N_m^-$ provide a complex scenario of their structure and energetics. Li and Zhao report that $N_5^-$ and $N_7^-$ are essentially van der Waals bound $N_3^-(N_2)_n$ species with binding energies of about 25 meV per $N_2$,[41] i.e. about twice the binding energy of $N_2$-$N_2$.[39] On the other hand, Law et al. characterize $N_7^-$ as a weakly bound $N_2N_5^-$ complex with $C_1$ symmetry.[42] Liu et al. report that the ground state structures of odd-numbered $N_m^-$ with $m \leq 11$ may be characterized as open chains with $C_{2v}$ symmetry [43] whereas $N_{15}^-$, which has $C_1$ symmetry, may be considered as a complex between a cyclic $N_5^-$ ($D_{5h}$) and staggered $N_{10}$ ($D_{2d}$).[44] The structure and energetics listed in their work suggests the presence of covalent bonding.

The anomaly at $n = 4$ in the ion abundance (Fig. 2a) suggests that the dissociation energies $D_n$ for the reaction

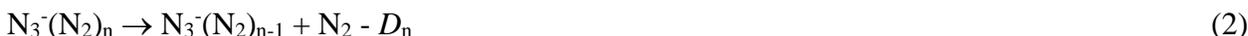
$N_3^-(N_2)_n \rightarrow N_3^-(N_2)_{n-1} + N_2 - D_n$ (2)

features a local maximum at $n = 4$. For a confirmation one would need to compute the energies of $N_3^-$ $(N_2)_n$ for, at the very least, $n = 3$, 4, and 5, or $N_9^-$, $N_{11}^-$, and $N_{13}^-$. Unfortunately, we are not aware of any calculations for $N_{13}^-$.

The anomaly in the ion abundance at $N_3^-(N_2)_{11}$ suggests that $D_{11}$ forms a local maximum. For confirmation one would need to compute the energies of $N_{23}^-$, $N_{25}^-$, and $N_{27}^-$ which are much larger than the largest anions considered so far in theoretical work. We note that, if one were to assume that $N^-$ forms the chromophore, the highly stable ion would be $N^-(N_2)_{12}$ which might have icosahedral structure. It is known for cationic clusters that the degree of charge localization, i.e. the nature of the chromophore, may depend on the size of the cluster.[45] However, the fact that free $N^-$ is not stable whereas free $N_3^-$ is highly stable with respect to dissociation as well as autodetachment makes the assumption of a $N^-$ chromophore rather implausible.

Let us now turn to a discussion of even-numbered $N_m^-$, $m \geq 4$. Electron scattering by $N_2$ molecules features a resonance near 2.3 eV, with an onset at 1.8 eV and vibrational structure.[19] The lifetime of the compound state is probably close to a single vibrational period. Angular distribution measurements confirms the designation of $^2\Pi_g$ for this first shape resonance.[19] The resonance also appears in experiments on $N_2$ doped van der Waals clusters[46] or water ice[47] where it provides for effective electron transfer and subsequent electron trapping in the host matrix.

Vostrikov and Dubov have measured absolute cross sections for anion formation by electron attachment to bare nitrogen clusters containing an estimated average of 125 to 520 $N_2$ molecules.[27] A resonance was observed near 2 eV. The authors concluded that electrons scatter inelastically via the $^2\Pi_g$ resonance and are then captured by the van der Waals cluster by polarization forces. Many other van der Waals clusters feature low energy resonances for non-dissociative attachment even though the corresponding monomers do not form stable anions.[36, 48-50]

In fact, direct capture of the incident electron causes another, weak resonance in the experiments by Vostrikov and Dubov at very low (<< 1 eV) energies. The authors did not record mass spectra but, barring contributions from impurities, product ions are probably of the form $(N_2)_n^-$ because dissociative attachment to $N_3^-(N_2)_n$ is not feasible at low energies, as discussed above.

We observe even-numbered $N_m^-$ ($m \geq 4$) just above the noise level, with an abundance of about 1.0 ± 0.3 % of adjacent odd-numbered $N_m^-$, in spectra that were recorded at electron energies between 10.5 and 50.5 eV. For nitrogen embedded in helium the $N_2^-$ ($^2\Pi_g$) resonance would be expected at 2.3 + 1.6 = 3.9 eV but we do not observe any even-numbered $N_m^-$ in a mass spectrum that was recorded with the electron energy being scanned from 1.4 to 13.7 eV. However, it is not warranted to draw any conclusions from this negative result. The ion yield in that spectrum was very weak, and we did not carefully search for other ions that might be resonantly formed at low energies.

### 4.2 Nitrogen cluster anions containing impurities

Looking at the mass spectrum in Fig. 1b we see several anions other than $N_m^-$. Here we offer some conjectures on their origin and likely chromophores. For a more comprehensive discussion one would need to consider the energetics of various ion molecule reactions which is beyond the scope of the present discussion. A measurement of ion yields *versus* electron energy would also provide valuable insight, but the statistics of the present data do not allow for such an analysis.

There are two reasons why large helium droplets are likely to contain impurities. First, the specified purity of the expanding helium gas is 99.9999 % which implies that, for an average droplet size of $10^5$, about 10 % of all droplets will contain an impurity. Second, large droplets have a large probability that they will collide with background gas on their way to the ionizer. As a result, there is a non-negligible probability that droplets contain $H_2O$, $O_2$, or hydrocarbon impurities.

Dissociative attachment to nitrogen clusters containing one or more of these impurities results in various anions[51] including a) $O^-(N_2)$, b) $O_2^-(N_2)_{n+1}$, c) $CN_3^-(N_2)_n$, and d) $C_2N_2^-(N_2)_n$. which we discuss very briefly:

a) The likely chromophore in $O^-(N_2)$, which probably results from $N_2$ clusters with an $O_2$ impurity, is $O^-$ which has a detachment energy of 1.44 eV.[40] An alternative ion, $N_2O^-$, is not stable with respect to autodetachment, and barely stable with respect to dissociation into $O^- + N_2$.[50]

b) $O_2^-(N_2)_{n+1}$ is nominally isobaric with $N_3^-(N_2)_nH_2O$, with a difference of $\Delta m = 0.024$ u. In the low mass range both ions are clearly identified. In the high mass range they are unresolved. A fit indicates that $N_3^-(N_2)_nH_2O$ dominates but both ions contribute. The electron affinity of $O_2$ is 0.44 eV.[40] The low electron affinity of NO (0.04 eV[40]) makes formation of $NO^-NO$ highly unlikely.

c) The carbon in $CN^-(N_2)_n$ may result from a hydrocarbon impurity although one may expect that the ion would retain some of the hydrogen. The electron affinity of the cyano radical is 3.86 eV.[40]

d) $C_2N_2^-(N_2)_n$. Cyanogen has aroused interest since the Cassini Plasma Spectrometer has detected numerous negative ions in Titan's upper atmosphere, including $C_2N_2^-$.[52] The ion has several stable conformers but their adiabatic electron affinities are modest, well below 1 eV.[53-54] $C_2^-(N_2)_{n+1}$ may be a better representation of this ion because the electron affinity of $C_2$ is 3.27 eV.[40]

**Conclusion**

We have presented the first experimental study of odd-numbered $N_m^-$ anions formed by electron attachment to nitrogen van der Waals clusters embedded in helium where $m$ covers a large range, from $m = 3$ to about 139. The threshold observed for formation of anions with $5 \leq m \leq 21$ agrees with the estimated thermodynamic threshold for $N_3^-$ formation if one takes into account the effect of the helium nanodroplet. We conclude that cluster ions are composed of an azide anion solvated in $N_2$. Theoretical work is called for to understand why $N_3^-(N_2)_4$ and $N_3^-(N_2)_{11}$ appear to be particularly stable.

**Acknowledgements**

This work was supported by the Austrian Science Fund, Wien (FWF Projects I978 and P26635).

**Figure Captions**

Figure 1
Fig. 1a displays a mass spectrum obtained by electron attachment to helium nanodroplets doped with $N_2$. The electron energy was repeatedly scanned in increments of 0.05 eV from 19.5 to 50.5 eV during the acquisition time of 10.3 hours. Thus, the spectrum represents an average over this range of electron energies. The dominant ion series consists of $N_3^-(N_2)_n$. Also marked are mass peaks due to $N_3^-(N_2)_n H_2O$ (with possible contributions from the nominally isobaric $O_2^-(N_2)_{n+1}$), and $C_2N_2^-(N_2)_n$. An exploded view of a narrow region of this spectrum is presented in Fig. 1b.

Figure 2
Ion abundances of $N_3^-(N_2)_n$ and $C_2N_2^-(N_2)_n$ *versus* size $n$ extracted from the mass spectrum in Fig. 1. Significant anomalies are labeled. Error bars in Fig. 2a are smaller than the symbol size.

Figure 3
Energy dependence of the cross section for formation of $N_3^-$ and $N_3^-(N_2)_n$, $1 \leq n \leq 9$ (Figs. 3a and 3b, respectively). The line in Fig. 3a indicates the result of a non-linear least squares fit of a Wannier-type power law with an exponent of 1.283, and taking into account the finite energy resolution (1.5 eV FWHM) of the electron beam. The threshold is $E_0 = 11.5 \pm 0.5$ eV. The threshold observed for $N_3^-(N_2)_n$ cluster ions is significantly lower.

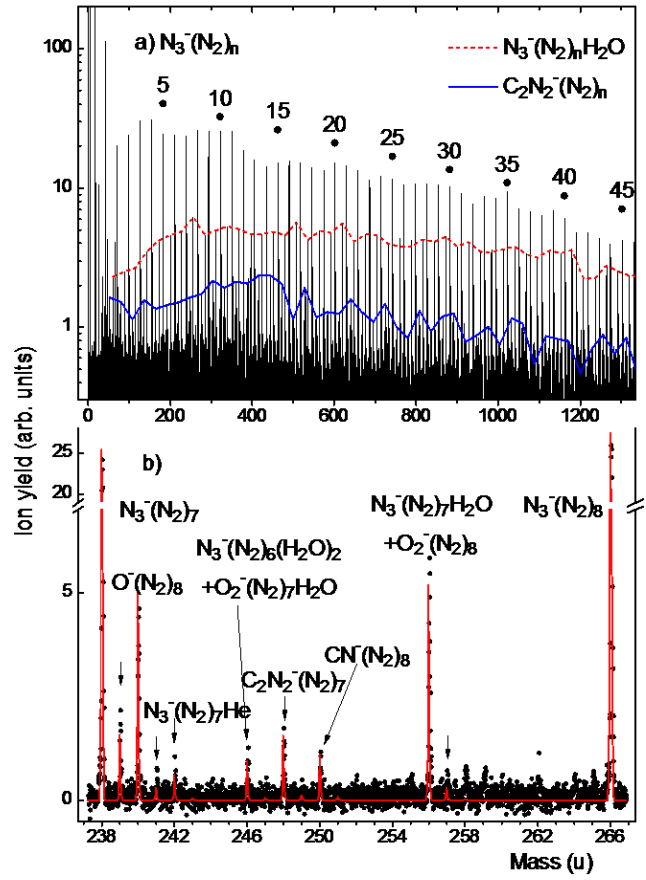

Fig. 1

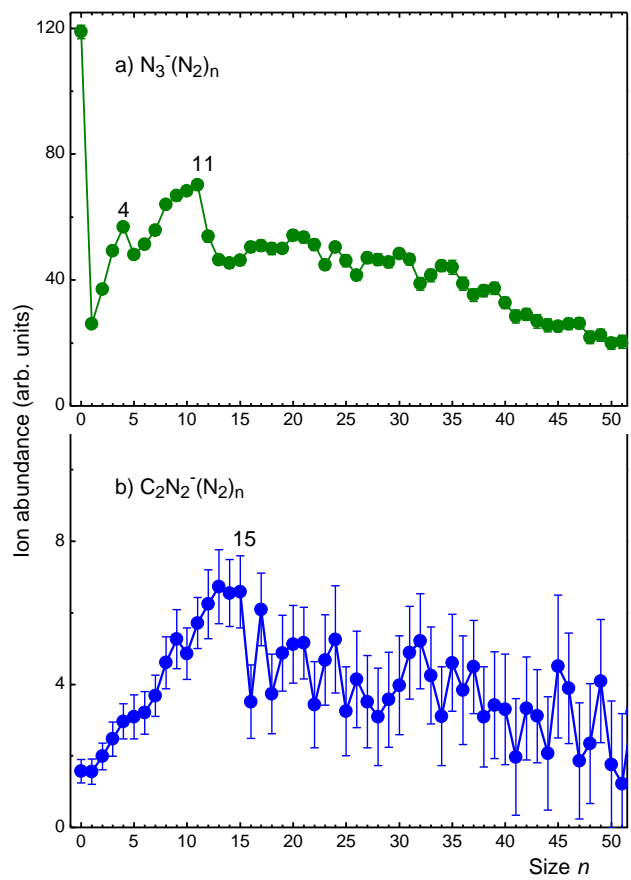

Fig. 2

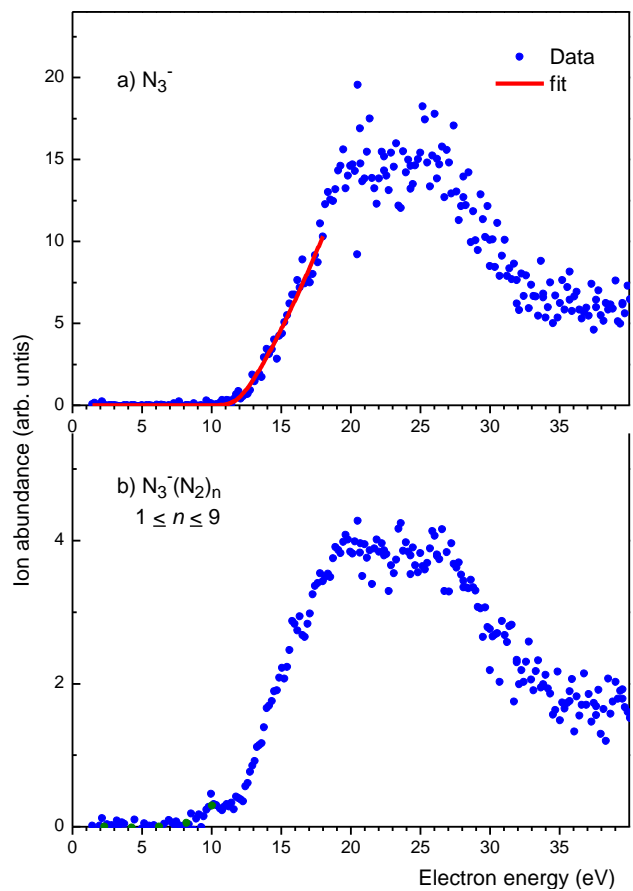

Fig. 3

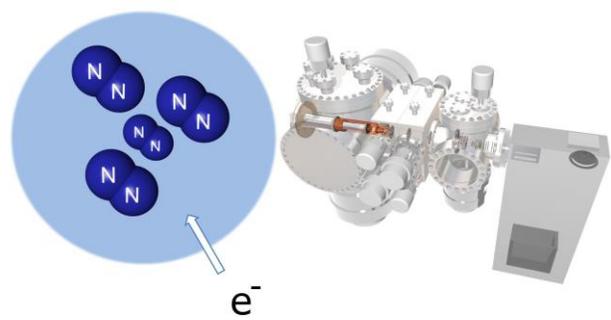

Table of Contents Graphic